\documentclass[conference,a4paper,dvipsnames]{IEEEtran}

\IEEEoverridecommandlockouts
\def\BibTeX{{\rm B\kern-.05em{\sc i\kern-.025em b}\kern-.08em
    T\kern-.1667em\lower.7ex\hbox{E}\kern-.125emX}}

\newcommand{\exportFigures}{false}
\usepackage{xcolor}

\usepackage{cite}
\usepackage{amsmath,amssymb,amsfonts,amsthm}
\usepackage{algorithmic}

\usepackage{graphicx}

\usepackage{textcomp}

\usepackage{url}
\usepackage{bm}

\usepackage[caption=false,font=footnotesize]{subfig}

\usepackage[normalem]{ulem}

\usepackage{mathtools, cuted}

\usepackage{pdfpages} %to include separate pdf pages

\usepackage{listings}

\usepackage{tikz}
\usetikzlibrary{angles}
\usetikzlibrary{arrows}
\usetikzlibrary{arrows.meta}
\usetikzlibrary{backgrounds}
\usetikzlibrary{calc}
\usetikzlibrary{decorations.pathmorphing}
\usetikzlibrary{decorations.text}
\usetikzlibrary{fit}
\usetikzlibrary{patterns}
\usetikzlibrary{petri}
\usetikzlibrary{plotmarks}
\usetikzlibrary{positioning}
\usetikzlibrary{quotes}
\usetikzlibrary{shapes.misc}
\usetikzlibrary{spy}

\usepackage{pgfplots}
\pgfplotsset{compat=newest}
\usepgfplotslibrary{patchplots}
\usepgfplotslibrary{units}
\usepgfplotslibrary{polar}

\usepackage[most]{tcolorbox}
\tcbuselibrary{breakable}
\tcbset{every box/.style={enhanced,breakable}}

\usepackage[ruled,vlined]{algorithm2e}
\SetKwRepeat{Do}{do}{while}
\usepackage{setspace} %for setstretch in algorithm environment (linespacing)

\usepackage{cancel}

\usepackage{ifthen}

\newcommand{\tikzpng}[2]{
\ifthenelse{#1=1}
{\centering\input{#2.tex}}
{\centering\includegraphics[width=\figurewidth]{#2.png}}
}

\ifthenelse{\equal{\exportFigures}{true}}
{
  \usepgfplotslibrary{external}\tikzexternalize[prefix=tikz/]
}
{}

\newcommand{\ticked}{$\text{\rlap{$\checkmark$}}\square$}
\newcommand{\unticked}{{$\square$}}
\newcommand{\tick}[1]{\ifthenelse{#1=1}{\ticked}{\unticked}}

\newcounter{assump}

\definecolor{tw}{RGB}{51,183,150}

\newcommand{\rmv}{\hspace*{-.3mm}}
\newcommand{\vm}[1]{\ensuremath{\bm{#1}}} %vector or matrix
\newcommand{\logl}[1]{\ensuremath{\text{log}\, #1 }}% logarithm (base 10) without brackets
\newcommand{\E}[1]{\ensuremath{\mathsf{E}[#1]}}% expectation operator
\DeclareMathOperator*{\argmax}{arg\,max}

\tcbset{colframe=black,colback=red!10,enhanced,breakable,sharp corners}
\tcbsetforeverylayer{shield externalize}

\newlength{\figureheight}
\newlength{\figurewidth}
\graphicspath{{./figures/}}

\usepackage[a4paper,left=1.43cm,right=1.43cm,top=1.8cm,bottom=4.43cm]{geometry}
\begin{document}

\title{Statistical Modeling of the Human Body as an Extended Antenna}

\author{\IEEEauthorblockN{
Thomas Wilding\IEEEauthorrefmark{4}\IEEEauthorrefmark{1},   % 1st author, 1st affiliations
Erik Leitinger\IEEEauthorrefmark{4},   % 2nd author, 2nd affiliations
Ulrich Muehlmann\IEEEauthorrefmark{3}      % 4th author, 4th affiliations
Klaus Witrisal\IEEEauthorrefmark{4}\IEEEauthorrefmark{1},    % 3rd author, 3rd affiliations
}                                     % ...
\thanks{This research was partly funded by the Austrian Research Promotion Agency (FFG) within the project UBSmart (project number: 859475). The financial support by the Christian Doppler Research Association, the Austrian Federal Ministry for Digital and Economic Affairs and the National Foundation for Research, Technology and Development is gratefully acknowledged.}
%\\
\IEEEauthorblockA{\IEEEauthorrefmark{4}% 1st affiliations
Graz University of Technology, Austria}
\IEEEauthorblockA{\IEEEauthorrefmark{3}% 3rd affiliations
NXP Semiconductors, Austria}
\IEEEauthorblockA{\IEEEauthorrefmark{1}% 2nd affiliations
Christian Doppler Laboratory for Location-aware Electronic Systems}
%\IEEEauthorblockA{\IEEEauthorrefmark{4}% 4th affiliations
%(Affiliation): dept. name of organization, name/acronyms of organization, City, Country,
% e-mail address*}  
%\IEEEauthorblockA{ \emph{thomas.wilding@tugraz.at} }
\IEEEauthorblockA{ \emph{thomas.wilding@tugraz.at} }
}

%\author{Thomas Wilding$^{1,2}$, Klaus Witrisal$^{1,2}$, Ulrich M\"uhlmann$^{3}$, 
%\thanks{This research was partly funded by the Austrian Research Promotion Agency (FFG) within the project UBSmart (project number: 859475). The financial support by the Christian Doppler Research Association, the Austrian Federal Ministry for Digital and Economic Affairs and the National Foundation for Research, Technology and Development is gratefully acknowledged.}
%\\
%\small{{$^1$Graz University of Technology, Austria}, {$^3$NXP Semiconductors, Austria}} \\
%\small{{$^2$Christian Doppler Laboratory for Location-aware Electronic Systems}} \\
%\small{email: thomas.wilding@tugraz.at} }

% \IEEEspecialpapernotice{``Its beauty is its simplicity.'' --- J. Kulmer}

\maketitle

\begin{abstract}
In this paper we investigate the possibility of modeling a single antenna alone and in close proximity to a physical object by means of discrete point source scatterers.
The scatter point model allows joint modeling of a physical antenna and the human body as a single extended object with direction dependent scattering coefficients for the scatter points.
We introduce the term extended antenna describing antenna and human body together. 
To investigate the identifiability of the model parameters we make use of ultrawideband channel measurements and accurate ground truth position and orientation measurements obtained with an optical tracking system.
By comparing measurements of the antenna attached directly to the user with measurements for the antenna without the user nearby, we show the shadowing and scattering effects of the human body and the antenna. 
\end{abstract}

\vskip0.5\baselineskip

%%%%%%%%%%%%%%%%%%%%%%%%%%%%%%%%%%%%%%%%%%%%%%%%%%%%%%%%%%%%%%%%%%%%%%%%%%%%%%%%
%%%%%%%%%%%%%%%%%%%%%%%%%%%%%%%%%%%%%%%%%%%%%%%%%%%%%%%%%%%%%%%%%%%%%%%%%%%%%%%%
%%%%%%%%%%%%%%%%%%%%%%%%%%%%%%%%%%%%%%%%%%%%%%%%%%%%%%%%%%%%%%%%%%%%%%%%%%%%%%%%
\section{Introduction}
\label{sec:introduction}

For positioning applications, an accurate model of the received signal is needed. This model should include any environment-related effects that can affect an estimator of position related signal parameters from the received signal. % infer position related parameters. 
Such effects can be specular multipath components (SMCs) and a diffuse multipath component (DMC) \cite{RichterPhD2005} or scattering objects such as mobile users \cite{SchmidhammerSensors2019} or trees, e.g., for GNSS signals \cite{SchubertTAP2013}. Especially blockage of the line-of-sight propagation path can introduce a strong bias in delay estimates, a survey of which is presented in \cite{guevencCST2009}. 
This is especially true in off-body communications, where the anchor or the agent are in close proximity to the user \cite{ambroziak2016,TurbicTAP2019}.
The effects of wearable antennas on narrowband signals were already investigated in \cite{mackowiak2012b,Mackowiak2013b} highlighting the strong shadowing effect introduced by the human body.
As shadowing alone is not sufficient to accurately model effects on wideband signals, it is of interest to develop suitable channel models incorporating both shadowing and dispersion in the delay-angle domain.

\begin{figure}[t!]\centering
%\vspace{1mm}
%\def\datapath{./figures/pgf/scattering_environment}
%\input{./figures/pgf/scattering_environment/scattering_environment.tex}
\includegraphics{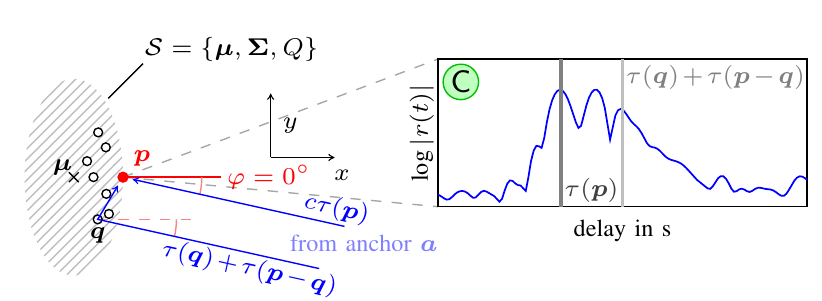}
\caption{Representation of an extended antenna $\mathcal{S}$ modeling a user in close proximity to the receive antenna at $\vm{p} = [x,y]^{\mathsf{T}}$ with a signal impinging from an anchor located at $\vm{a} = [x_a, y_a]^\mathsf{T}$. Scattering points are indicated as $\vm{q} = [x_q, y_q]^{\mathsf{T}}$. An exemplary received signal $r(t)$ for the receiver at on-body position $\mathsf{C}$ is shown for visualization purpose.}\label{fig:scattering_environment}
\end{figure}

In \cite{WildingPIMRC2020} we have shown that a user equipped with an agent can be modeled as an extended object (EO) using a measurement based, empirically chosen scattering point distribution and a gain function.
In this paper we investigate the possibility of calibrating the statistical model and show that the joint description of scattering object and non-ideal antenna is implicitly possible with the proposed model from \cite{WildingPIMRC2020}.
We model agent and user as one EO under the term extended antenna (EA), adding components via point source scatterers jointly modeling the dispersive effects of antenna and user. 
The general setup under investigation is shown in Fig.~\ref{fig:scattering_environment}, with the EA denoted by $\mathcal{S}$ and the reference direction for the angle of arrival (AOA) $\varphi$ included as well. 

For estimation and tracking of the shape of EOs, a large number of algorithms exist \cite{KochTAES2008,GranstroemArxiv2016,LundgrenTSP2016,YangTSP2019}, which usually require a large number of (spatial) measurements (e.g., obtained using LIDAR sensors \cite{GranstroemArxiv2016}) to accurately estimate the object's state.
As highly accurate spatial measurements using radio signals require large antenna arrays and a large measurement bandwidth, state of the art algorithms are not directly applicable. 
This paper shows that a suitable measurement setup allows estimation of scattering points to calibrate the model. \IEEEpubidadjcol

The rest of the paper is structured as follows: Section~\ref{sec:system_model} describes the system and signal model, Section~\ref{sec:validation} introduces the measurement campaign and the estimation procedure. Results are shown in Section~\ref{sec:results} and Section~\ref{sec:conclusion} concludes the paper.

%%%%%%%%%%%%%%%%%%%%%%%%%%%%%%%%%%%%%%%%%%%%%%%%%%%%%%%%%%%%%%%%%%%%%%%%%%%%%%%%
%%%%%%%%%%%%%%%%%%%%%%%%%%%%%%%%%%%%%%%%%%%%%%%%%%%%%%%%%%%%%%%%%%%%%%%%%%%%%%%%
%%%%%%%%%%%%%%%%%%%%%%%%%%%%%%%%%%%%%%%%%%%%%%%%%%%%%%%%%%%%%%%%%%%%%%%%%%%%%%%%
\section{System Model and Signal Model}
\label{sec:system_model}
%\input{./input_files/system_model.tex}

% This section describes the received signal model and introduces point processes as an intuitive tool for joint modeling of channel and one or multiple antennas. 
% Based on our investigation in \cite{WildingPIMRC2020} we reformulate the model with the aim of direct positioning in mind, setting the framework for the following evaluation. 

\subsection{System Model}
We consider a single anchor at known position $\vm{a} = [x_a,y_a]^\mathsf{T} \in \mathbb{R}^2$ and an agent at position $\vm{p}=[x,y]^\mathsf{T}\in \mathbb{R}^2$, each equipped with a single antenna (see Fig.~\ref{fig:scattering_environment}). 
The EA $\mathcal{S} = \{\vm{\mu},\vm{\Sigma},Q\}$ is described by mean position $\vm{\mu} = [x_\mathcal{S}, y_\mathcal{S}]^{\mathsf{T}}\in\mathbb{R}^2$ with the matrix $\vm{\Sigma}\in\mathbb{R}^{2\times2}$ denoting the shape in terms of a covariance matrix. 
Scattering points $\vm{q}_j=[x_j,y_j]^{\mathsf{T}}\in\mathbb{R}^2$ belonging to the EA $\mathcal{S}$ are modeled as a marked Poisson point process (MPPP) denoted by $Q = \{(\vm{q}_1,\beta_1), (\vm{q}_2,\beta_2), \dots \} = \{(\vm{q}_j,\beta_j)\}$ where $j$ is an index solely used to distinguish points and $\beta_j\in\mathbb{C}$ is the complex mark of the corresponding scattering point $\vm{q}_j$. %modeled using a suitable distribution and $N(Q)$ represents the average number of points in the MPPP. 
Per definition of an MPPP, the number of points in $Q$ is Poisson distributed with mean $N(Q)$, with the points spatially distribution in the EA shape according to an intensity function $\rho(\vm{q})$. %(homogeneous or in-homogeneous).
A thorough treatment of point process theory can be found in \cite{illian2008}.
The extension of the model to 3-dimensions is straightforward.

\subsection{Signal Model}
The signal at the agent is modeled to consist of a deterministic and a stochastic part. The former contains $K$ specular multipath components (SMCs), e.g., modeled by an image source model \cite{KulmerTWC2018}. The latter contains a scattered component (SC) and a diffuse multipath component (DMC) modeling the environment randomness. The SC is modeled by the MPPP $Q$ described above and the DMC by a random process $\nu(\tau)$ with a suitable delay-angle power spectral density, see, e.g., \cite{RichterPhD2005,hinteregger2016}.
We are only interested in effects on the line-of-sight (LOS) path between anchor and agent, i.e., without loss of generality we set $K=1$ and neglect the DMC.
% Due to the linearity of the system, the extension to SMCs is straightforward.
% Similarly, also the separation of the channel into a deterministic and a stochastic component is already 
% Similar to \cite{SchubertTAP2013,WildingPIMRC2020}, we separate the spreading function $h(\tau)$ into a deterministic and a stochastic component
% \begin{align}
% h(\tau) &= h_\mathrm{d}(\tau) + h_\mathrm{s}(\tau) \label{eq:channel}
% \end{align}
% with the deterministic component $h_\mathrm{d}(\tau)$ representing SMCs and the stochastic component $h_\mathrm{s}(\tau)$ the DMC and otherwise scattered components, e.g., by EOs.
% Due to linearity, we directly formulate the received signal as the output of a system with spreading function from \eqref{eq:channel}.
This gives the radio channel as
\begin{align}
h(\tau;\vm{p}) \rmv=\rmv \alpha \delta(\tau\rmv-\rmv\tau(\vm{p})) \rmv+\rmv \sum_{j} \beta_{j}\delta(\tau\rmv-\rmv \tau(\vm{q}_j)\rmv-\rmv\tau(\vm{p}\rmv-\rmv\vm{q}_j))\nonumber
\end{align}
where $\alpha\in\mathbb{C}$ is the complex LOS amplitude and $\beta_{j}\in\mathbb{C}$ the complex scattering coefficient of scattering point $\vm{q}_j$. The path delays are $\tau(\vm{p}) = \frac{1}{c}\|\vm{p}-\vm{a}\|$ where $\|\cdot\|$ denotes the vector norm and $c$ is the speed of light.

The signal $r(t)$ received at the agent when transmitting a baseband pulse $s(t)$ at carrier frequency $f_c$ is given by
\begin{align}
r(t) &= \int s(t-\tau)e^{-i2\pi f_c \tau}h(\tau;\vm{p}) d\tau + n(t) \\
&= \alpha s(t;\vm{p}) + \sum_{j} \beta_{j}s(t;\vm{p},\vm{q}_j) + n(t) \label{eq:rt}
\end{align}
where $s(t;\vm{p}) = s(t-\tau(\vm{p}))e^{-i2\pi f_c \tau(\vm{p})}$ is the pulse received at position $\vm{p}$ and 
$s(t;\vm{p},\vm{q}) = s(t-\tau(\vm{p}-\vm{q});\vm{q})e^{-i2\pi f_c \tau(\vm{q}-\vm{q})}$
%\begin{align}
%% s(t;\vm{p},\vm{q}) &= s(t-\tau(\vm{q})-\tau(\vm{p}-\vm{q}))e^{-i2\pi f_c \tau(\vm{q})}e^{-i2\pi f_c \tau(\vm{p}-\vm{q})} \\
%s(t;\vm{p},\vm{q}) &= s(t-\tau(\vm{p}-\vm{q});\vm{q})e^{-i2\pi f_c \tau(\vm{q}-\vm{q})}
%\end{align}
the pulse received at $\vm{p}$ after scattering at $\vm{q}$.
The measurement noise $n(t)$ is modeled as zero-mean complex additive white Gaussian noise (AWGN) with double-sided power spectral density $\frac{N_0}{2}$.
Synchronous sampling of \eqref{eq:rt} with frequency $f_s = \frac{1}{T_s}$ over an observation interval $T$ and stacking the samples in the vector $\vm{r} \triangleq [r(0), r(T_s),\dots,r((N-1)T_s)]^\mathsf{T} \in\mathbb{C}^N$ and $N = T/T_s$ yields the discrete-time signal model  %We obtain the discrete-time signal model as
\begin{align}
\vm{r} &=\rmv \alpha\vm{s}(\vm{p}) \rmv+\rmv \sum_{j}\beta_{j} \vm{s}(\vm{p},\vm{q}_j) \rmv+\rmv \vm{n} ~\in\mathbb{C}^{N} \label{eq:model} \\
&\triangleq\rmv \alpha\vm{s}(\vm{p}) \rmv+\rmv  \vm{S}(\vm{p},\vm{Q})\vm{\beta} \rmv+\rmv \vm{n} \label{eq:inference}%\\
% &= \vm{S}(\vm{p},\vm{q})\vm{a} + \vm{n}
\end{align}
where $\vm{\beta}=[\beta_1,\dots, \beta_{J}]^\mathsf{T}\in\mathbb{C}^{J}$ and $\vm{S}(\vm{p},\vm{Q})=[\vm{s}(\vm{p},\vm{q}_1),\dots,\vm{s}(\vm{p},\vm{q}_{J})]\in\mathbb{R}^{N\times J}$ where $J$ denotes the number of components for a specific realization of $Q$. Scattering point positions are combined into $\vm{Q}=[\vm{q}_1^\mathsf{T},\dots,\vm{q}_{J}^\mathsf{T}]$ and the columns of $\vm{S}(\vm{p},\vm{Q})$ are defined as $\vm{s}(\vm{p}) \triangleq [s(0;\vm{p}),\dots, s((N-1)T_s;\vm{p})]^{\mathsf{T}}\in \mathbb{C}^{N}$ and $\vm{s}(\vm{p},\vm{q}) \triangleq [s(0;\vm{p},\vm{q}),\dots, s((N-1)T_s;\vm{p},\vm{q})]^{\mathsf{T}}\in \mathbb{C}^{N}$.
%\begin{align}
%\vm{s}(\vm{p}) &\triangleq [s(0;\vm{p}),\dots, s((N-1)T_s;\vm{p})]^{\mathsf{T}}\in \mathbb{C}^{N} \label{eq:s_vec_p}\\
%\vm{s}(\vm{p},\vm{q}) &\triangleq [s(0;\vm{p},\vm{q}),\dots, s((N-1)T_s;\vm{p},\vm{q})]^{\mathsf{T}}\in \mathbb{C}^{N}\label{eq:s_vec_pq}
%\end{align}
The vector $\vm{n}$ contains the AWGN noise samples with covariance matrix $\vm{C} \triangleq \E{\vm{n}\vm{n}^\mathsf{H}} = \sigma_w^2 \vm{I}_N$, $\sigma_w^2=\frac{N_0}{T_s}$ and $\vm{I}_N$ is an $N\times N$ identity matrix.

The model in \eqref{eq:inference} can be interpreted as a standard multipath model with $J+1$ components: the LOS path directly arriving at the agent $\vm{p}$ from the anchor $\vm{a}$ and $J$ components that are paths scattered at locations $\vm{q}_j$.
The according likelihood function of the model is
\begin{align}\label{eq:likelihood}
% f(\vm{r};\vm{p},\alpha,\vm{\theta},\vm{\beta}) = \frac{e^{-(\vm{r} - \vm{S}(\vm{p},\vm{\theta})\vm{b})^\mathsf{H} \vm{C}^{-1}(\vm{r} - \alpha\vm{s}(\vm{p}) - \vm{S}(\vm{p},\vm{\theta})\vm{\beta})}}{\pi^{N} \det(\vm{C})}
f(\vm{r};\vm{p},\alpha,\vm{Q},\vm{\beta}) \propto \mathrm{exp}(-\|\vm{r} - \alpha\vm{s}(\vm{p}) - \vm{S}(\vm{p},\vm{Q})\vm{\beta}\|^2).
\end{align}

Due to the large number of model parameters, estimation of all parameters might not be possible, depending on the measurement setup. 
The following section describes a measurement procedure to allow calibration of $\vm{\beta}$ and $\vm{Q}$ by forming a synthetic array assuming a known agent position $\vm{p}$.
 
%To overcome the identifiability issues of the model parameters, one needs to use a suitable measurement setup which allows combining multiple snapshots to formulate an artificial estimator for model calibration.
%The artificial estimation procedure for scattering point positions $\vm{Q}$ and amplitudes $\vm{\beta}$ is formulated in the next section.

% As a quick estimate of the mark distribution we use the likelihood of the signal in a grid around the true position of the EO. Using multiple measurements in combination with the highly accurate ground truth positions obtained from the optical tracking system and the large bandwidth of the channel sounder results in a high spatial resolution.

%%%%%%%%%%%%%%%%%%%%%%%%%%%%%%%%%%%%%%%%%%%%%%%%%%%%%%%%%%%%%%%%%%%%%%%%%%%%%%%%
%%%%%%%%%%%%%%%%%%%%%%%%%%%%%%%%%%%%%%%%%%%%%%%%%%%%%%%%%%%%%%%%%%%%%%%%%%%%%%%%
%%%%%%%%%%%%%%%%%%%%%%%%%%%%%%%%%%%%%%%%%%%%%%%%%%%%%%%%%%%%%%%%%%%%%%%%%%%%%%%%
\section{Measurement Setup and Model Calibration}
\label{sec:validation}

To calibrate the model ultrawideband measurements were performed alongside ground truth position and orientation measurements of the agent and parts of the user, obtained using an optical tracking system (OTS) with position and orientation accuracy below {\color{black}$1\,\mathrm{mm}$ and $1^\circ$} respectively.
This setup allows using a synthetic array of anchors to formulate a non-coherent estimation procedure of scattering points and coefficients.

\subsection{Measurement Setup}\label{sec:measurements}
\begin{figure}[t!]
\centering
% \subfloat[channel sounder]
% {\includegraphics[width=0.2\textwidth]{./figures/png/channel_sounder_orig.png}\label{fig:hardware:channel_sounder}}\hspace{10mm}
% \subfloat[anchor $\vm{a}$]
% {\includegraphics[width=0.17\textwidth]{./figures/png/anchor.png}\label{fig:hardware:anchor}}%\hspace{0mm}
\subfloat[mobile at $\mathsf{R}$]
{\includegraphics[width=0.12\textwidth]{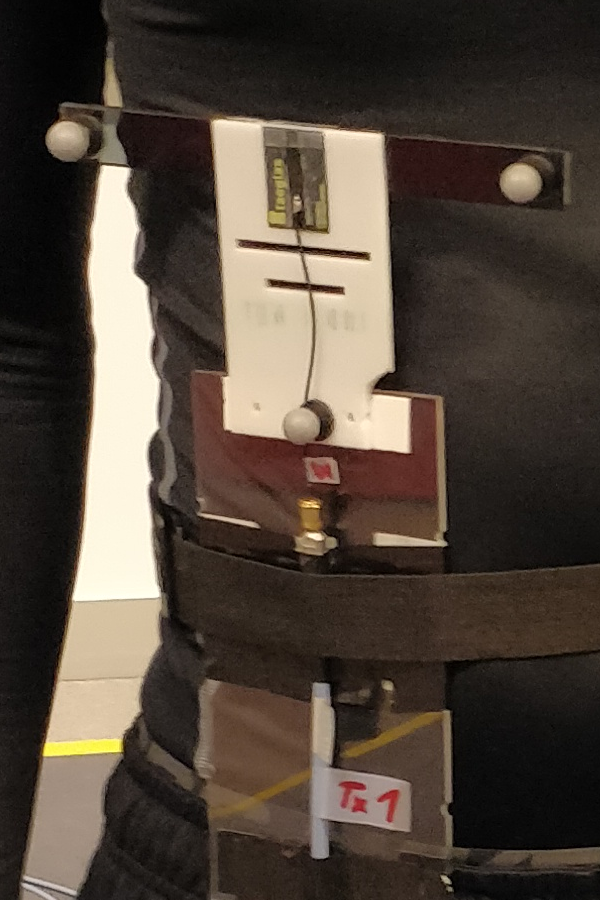}
\label{fig:hardware:mobile}}%\hspace{10mm} %or [width=0.146\textwidth,height=0.23\textwidth]{./figures/mobile.png}
\subfloat[floorplan]{
\includegraphics[width=0.22\textwidth]{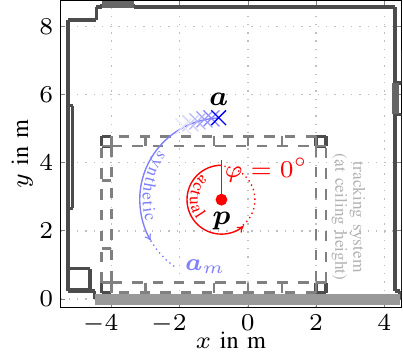}
\label{fig:hardware:floorplan}
}

\caption{The agent at on-body position $\mathsf{R}$ with reflective markers for optical tracking system is shown in \ref{fig:hardware:mobile}. The floorplan shows the position $\vm{a}$ of the anchor (blue x) and the approximate agent $\vm{p}$ (red dot).% \eqref{pgf:p}. %The mounting beams at ceiling height of the tracking system cameras is indicated as \ref{pgf:track_mounting}.
}\label{fig:hardware}
\end{figure}

The radio measurements were performed with a correlative channel sounder (CS) covering a frequency range of $f=3.8-10.2 \,\mathrm{GHz}$ with $f_c=6.95\,\mathrm{GHz}$ allowing fully coherent measurements. 
The measurement system was calibrated, including CS, cables and connectors (but not the antennas).
The antennas used were a {taoglas FXUWB10} patch antenna as agent (see Fig.~\ref{fig:hardware:mobile}) and a dipole antenna \cite[App.~B.3]{KrallPhD2008} as anchor, both suitable for the measurement frequency range.
The dipole antenna has a sufficiently flat pattern in the horizontal plane with strong attenuation towards floor and ceiling, reducing floor and ceiling reflections. Both devices are positioned at similar height.

\begin{figure}[t!]
\subfloat[body and ref.]{\centering
{\scalebox{0.825}{%\input{./figures/pgf/human_front_reference.tex}
\includegraphics{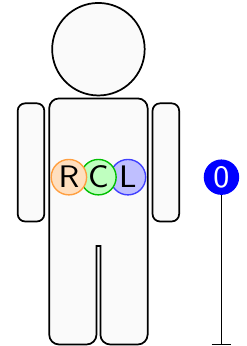}\label{fig:setup:body}}}\hspace{1mm}}
\subfloat[$\mathsf{C}$]{\centering
\includegraphics[height=3cm]{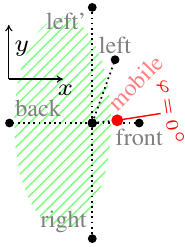}\label{fig:setup:L}\hspace{1mm}}
\subfloat[$\mathsf{L}$]{\centering
\includegraphics[height=3cm]{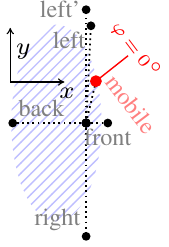}\label{fig:setup:C}\hspace{1mm}}
\subfloat[$\mathsf{R}$]{\centering
\includegraphics[height=3cm]{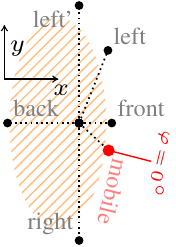}\label{fig:setup:R}}

\caption{Used on-body positions of the mobile device are shown in \ref{fig:setup:body}. Reference coordinate system for the AOA $\varphi$ (positive AOAs are defined counter clockwise with the mobile position $\vm{p}$ (red dot) as reference point) and the tracked body part positions of upper back (\textit{back}), right lower arm and left lower arm (\textit{right} and \textit{left}) and a cap (\textit{front}) are shown in \ref{fig:setup:L}-\ref{fig:setup:R}. The dashed areas give a rough outline of the user. %The point \textit{left'} is the point \textit{right} mirrored along the axis \textit{back}--\textit{front}.
} 
\label{fig:setup}
\end{figure}

The anchor position was fixed throughout the measurements, with the agent positioned at various on-body positions (see Fig.~\ref{fig:setup:body}) on a user performing a full rotation on the spot with a static posture and static on-body position of the agent, while recording $M = 200$ channel responses. %\footnote{It is assumed that slight variations in AOD from $\varphi_1 = 0\,\mathrm{rad}$ are negligible as the dipole antennas are omnidirectional in the horizontal plane.}
%During each measurement run, the on-body position of the agent was fixed, allowing to characterize the position separately.
An overview of the on-body positions is shown in Fig.~\ref{fig:setup:body} alongside the coordinate system used in the evaluation.
The floorplan of the measurement setup is shown in Fig.~\ref{fig:hardware:floorplan}, indicating the user position and rotation direction. 
Using the OTS data it is possible to reverse the setup and synthetically generate a static agent and multiple anchors around the agent. 
This procedure allows to jointly use all $M=200$ measurements obtained during a rotation of the user, simulating a synthetic array by fixing the agent device (and thus the EA) and moving the anchor at (synthetic) positions $\vm{a}_m$ on a circle centered around the agent (see Fig.~\ref{fig:hardware:floorplan}).
Note that this is only valid for investigation of the LOS.
% Ground truth orientation and position data of anchor, agent and body parts were obtained using an optical tracking system (OTS).

\subsection{Model Calibration}
\label{sec:estimation}
To calibrate the model parameters we make use of the available ground truth anchor and agent positions and apply a maximum likelihood (ML) estimator to directly estimate scatter point positions. 
The underlying assumption is that due to the static posture and on body position, only one realization of the EA is observed during the $M$ measurements.
Assuming furthermore that all measurements $\vm{r}_m$ are independent with identical noise variance\footnote{This is a valid assumption as the measurements were performed in a short time frame with the measurement device in continuous use.} $\sigma_w^2$, allows using \eqref{eq:likelihood} to write the joint likelihood of all measurements as
\begin{align}\label{eq:joint_likelihood}
f(\underline{\vm{r}};\vm{p},\underline{\vm{\alpha}},\vm{Q},\underline{\vm{\beta}}) &=\rmv\rmv \prod_{m=1}^{M} f(\vm{r}_m;\vm{p},\alpha_m,\vm{Q},\vm{\beta}_m) %\\
% &\propto\rmv \mathrm{exp}\Bigl(\sum_{m=1}^M \rmv\rmv-\|\vm{r} \rmv-\rmv \alpha\vm{s}(\vm{p}) \rmv-\rmv \vm{S}(\vm{p},\vm{\theta})\vm{\beta}\|^2\Bigr)
\end{align}
where $\underline{\vm{r}} = [\vm{r}_1^\mathsf{T}, \dots,\vm{r}_M^\mathsf{T}]^\mathsf{T}$, $\underline{\vm{\alpha}}=[\alpha_1,\dots, \alpha_M]^\mathsf{T}$, $\underline{\vm{\beta}} = [\vm{\beta}_1^\mathsf{T}, \dots,\vm{\beta}_M^\mathsf{T}]^\mathsf{T}$ are the stacked signals, LOS amplitudes and scattering coefficients for all $M$ measurement.
Based on \eqref{eq:joint_likelihood} the resulting ML estimator is obtained as
\begin{align}\label{eq:mle}
\underline{\hat{\vm{\alpha}}}, \underline{\hat{\vm{\beta}}}, \hat{\vm{Q}} = \argmax_{\vm{\alpha},\vm{\beta},\vm{Q}} \prod_{m=1}^{M} f(\vm{r}_m;\vm{p},\alpha_m,\vm{Q},\vm{\beta}_m).
\end{align}

% Under the assumption of static on body position and static user posture, the scattering point positions in $\vm{Q}$ and the agent position $\vm{p}$ are common parameters to all likelihoods in \eqref{eq:joint_likelihood}. 
%Taking the logarithm of \eqref{eq:joint_likelihood} finally yields the log likelihood  
%\begin{align}
%\logl f(\underline{\vm{r}};\vm{p},\underline{\vm{\alpha}},\rmv\vm{Q},\rmv\underline{\vm{\beta}}) &\rmv\propto\rmv\rmv\rmv \sum_{m=1}^M \rmv\rmv-\|\vm{r}_m \rmv-\rmv \alpha_m\vm{s}(\vm{p}) \rmv-\rmv \vm{S}(\vm{p},\rmv\vm{Q})\vm{\beta}_m\|^2\label{eq:full_lhf}
%\end{align}
%to maximize.

To optimize \eqref{eq:mle}, the LOS amplitudes and scattering coefficients are estimated separately for each measurement $m$ as linear least squares (LS) estimates \cite{golub1973} denoted by $\hat{\underline{\vm{\alpha}}}$ and $\hat{\underline{\vm{\beta}}}$ using the known ground truth agent position $\vm{p}_t$ available from the OTS.\footnote{Anchor positions $\vm{a}_m$ are also obtained by the OTS and assumed known in the signal model as usually the case in positioning applications.}
These estimates can be interpreted as non-parametric estimates of the user shadowing and agent antenna pattern, modeling the full gain pattern of the EA. 
Delay dispersion of the user and the agent antenna is in turn modeled by the scattering points. 
To find estimates of scattering points $\hat{\vm{Q}}$ we employ a simple procedure which greedily maximizes \eqref{eq:joint_likelihood} by adding scattering points consecutively (i.e., without joint optimization) as long as the likelihood value can be increased. 
The chosen stopping criterion is by no means optimal and was chosen heuristically.
%, resulting in $J\leq9$ estimated scattering points in the investigated measurement setups $\mathsf{0}$, $\mathsf{L}$, $\mathsf{C}$ and $\mathsf{R}$ (see Fig.~\ref{fig:setup:body}). 
%Furthermore, sophisticated algorithms can be employed, but as no ground truth was available, a thorough performance characterization would not be possible.
Even though more sophisticated algorithms could be used, the effect of the human body and the antenna alone can be compared nonetheless as we apply the same algorithm to measurements of the agent with and without user, with (possible) estimation artifacts arising in both cases.
An overview of the described iterative ML procedure is given in Algorithm~\ref{alg:mle}.

\begin{algorithm}[t!]\label{alg:mle}\footnotesize%\small%\setstretch{1}
\textbf{Initialization:}
\begin{itemize}\setlength\itemsep{0.25mm}
\item $\hat{\underline{\vm{\alpha}}}=[\hat{\alpha}_1,\dots,\hat{\alpha}_M]^{\mathsf{T}} ~\leftarrow~ \hat{\alpha}_m = \frac{\vm{s}(\vm{p}_t)^{\mathsf{H}}\vm{r}_m}{\vm{s}(\vm{p}_t)^{\mathsf{H}}\vm{s}(\vm{p}_t)}~~\forall\,m$\;
\item $j = 0$\;
\end{itemize}
\textbf{Iterations:} \\
\Do{$\mathrm{log} f(\underline{\vm{r}};\vm{p}_{t},\underline{\hat{\vm{\alpha}}},\rmv\hat{\vm{Q}}^{(j)}\rmv\rmv,\hat{\underline{\vm{\beta}}}^{(j)})\rmv\rmv<\rmv\mathrm{log} f(\underline{\vm{r}};\vm{p}_{t},\underline{\hat{\vm{\alpha}}},\rmv\hat{\vm{Q}}^{(j-1)}\rmv\rmv,\hat{\underline{\vm{\beta}}}^{(j-1)})$}
{
$j~\leftarrow~j + 1$\;
find $\hat{\vm{q}}_{j} = \argmax\limits_{\vm{q}} \sum_{m=1}^M \logl f(\vm{r}_{m,j};\vm{p}_{t},\hat{\alpha}_m,\vm{q},\hat{\beta}_{m,j})$\;
~~~using $\vm{r}_{m,j} = \vm{r}_m - \vm{s}(\vm{p}_t)\hat{\alpha}_m - \sum_{l=1}^{j-1}\vm{s}(\vm{p}_{t},\hat{\vm{q}}_l)\hat{\beta}_{m,l}$\;
% ~~~using $\hat{\beta}_{m,j} = \frac{\vm{s}(\vm{p}_{t},\hat{\vm{q}}_k)^{\mathsf{H}}\vm{r}_{m,j}}{\|\vm{s}(\vm{p}_{t},\hat{\vm{q}}_j)\|^2}$\;
~~~using $\hat{\beta}_{m,j} = \|\vm{s}(\vm{p}_{t},\hat{\vm{q}}_k)\|^{-2}\vm{s}(\vm{p}_{t},\hat{\vm{q}}_k)^{\mathsf{H}}\vm{r}_{m,j}$\;
$\underline{\hat{\vm{\beta}}}^{(j)}~\leftarrow~\{\hat{\beta}_{1,1},\dots,\hat{\beta}_{M,j-1}\}~\leftarrow~\hat{\beta}_{m,j}$\;
${\hat{\vm{Q}}}^{(j)}~\leftarrow~\{\hat{\vm{q}}_1,\dots,\hat{\vm{q}}_{j-1}\}~\leftarrow~\hat{\vm{q}}_{j}$\;
}
\caption{iterative estimation procedure}
\end{algorithm}

%%%%%%%%%%%%%%%%%%%%%%%%%%%%%%%%%%%%%%%%%%%%%%%%%%%%%%%%%%%%%%%%%%%%%%%%%%%%%%%%
%%%%%%%%%%%%%%%%%%%%%%%%%%%%%%%%%%%%%%%%%%%%%%%%%%%%%%%%%%%%%%%%%%%%%%%%%%%%%%%%
%%%%%%%%%%%%%%%%%%%%%%%%%%%%%%%%%%%%%%%%%%%%%%%%%%%%%%%%%%%%%%%%%%%%%%%%%%%%%%%%
\section{Results}
\label{sec:results}
%\input{./input_files/results.tex}

\begin{figure}[t!]\centering
\vspace{3mm}

\subfloat[$\mathsf{L}$ (left upper torso)]{\hspace{0mm}\centering
\includegraphics{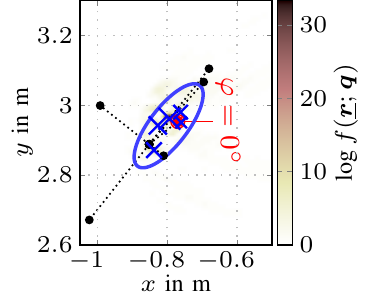}\label{fig:scatter_points:L}\hspace{0mm}}
\subfloat[$\mathsf{C}$ (center upper torso)]{\centering
\includegraphics{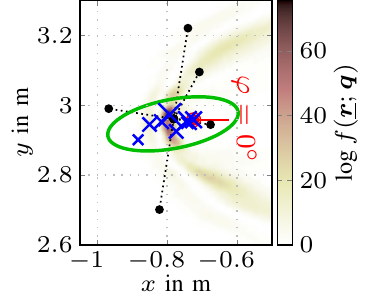}\label{fig:scatter_points:C}}%\hspace{-3mm}}

\subfloat[$\mathsf{R}$ (right upper torso)]{\hspace{0mm}\centering
\includegraphics{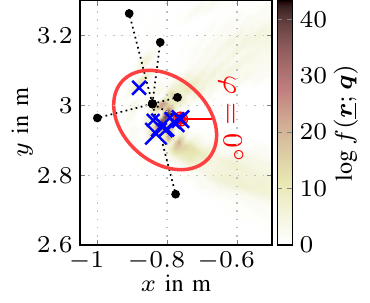}\label{fig:scatter_points:R}\hspace{0mm}}
\subfloat[$\mathsf{0}$ (reference)]{\centering
\includegraphics{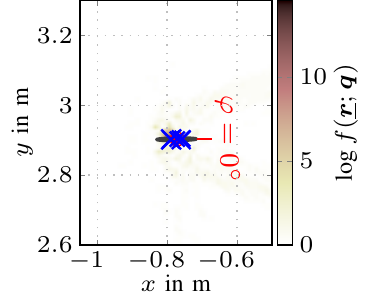}\label{fig:scatter_points:0}}

\caption{Estimated scatter points using the MLE algorithm described in Algorithm~\ref{alg:mle} with signals recorded at the different on-body positions and the agent alone as shown in Fig.~\ref{fig:setup:body}. The covariance ellipses (2-fold enlarged) of the scattering points are shown as EO shape estimate.}
\label{fig:scatter_points}
\end{figure}

%\ref{pgf:A1} and \ref{pgf:scatterer} have the same mark, but the first looks nicer
Fig.~\ref{fig:scatter_points} shows the scattering point estimates $\hat{\vm{Q}}$ (denoted by blue x) obtained using Algorithm~\ref{alg:mle} for the agent at different on-body positions (see Fig.~\ref{fig:setup:body}).
The size of the scattering point visualizes the respective average scattering coefficient magnitude $\bar{\beta}_j$ of the $j$th scattering point, defined as 
\begin{align}
\bar{\beta}_j = \frac{1}{M}\sum_{m=1}^{M} |\hat{\beta}_{m,j}|
\end{align}
with $\hat{\beta}_{m,j}$ being the estimated scattering coefficient of measurement $m$ and scattering point $j$.
An identical definition is used for the average LOS magnitude $\bar{\alpha} = \frac{1}{M} \sum_{m=1}^M |\hat{\alpha}_m|$. %$m$ denotes the measurement index. 
We include the scattering point covariance matrices for each on-body position computed as 
\begin{align*}
\hat{\vm{\Sigma}} = \frac{1}{J}\sum_{j=1}^J (\vm{q}_j-\hat{\vm{\mu}})(\vm{q}_j-\hat{\vm{\mu}})^{\mathsf{T}}~~\text{with}~~\hat{\vm{\mu}} = \frac{1}{J}\sum_{j=1}^J \vm{q}_j
\end{align*}
as simple estimates of the EA shape and position parameters respectively. 
Comparing the results at different on-body agent positions with the reference agent shows a strong increase in the spatial spread of the scattering points, attributable to the human body. 
The orientation of the covariance ellipses shows a slight correspondence to the employed on body position, with a larger spread towards the sides when at an off-center on-body position ($\mathsf{L}$ and $\mathsf{R}$) and larger spread towards the the body for the roughly symmetric position $\mathsf{C}$.
The fact that also the reference agent exhibits a non-zero number of estimated scattering points, but with much lower spatial spread, supports the proposed joint model of agent and user as an EA.

% 3cm -> 2.8cm -> 2.5cm

% \begin{figure*}[t!]%\centering
% \subfloat[$\mathsf{C}$ (center upper torso)]{\centering
% \setlength{\figureheight}{3cm}
% \setlength{\figurewidth}{3cm}
% \def\datapath{./figures/pgf/scatter_points_v3/m107/m107_llhf_K9}
% \input{./figures/pgf/scatter_points_v3/m107/m107_llhf_K9/m107_llhf_K9.tex}}%\hspace{-3mm}}
% \subfloat[$\mathsf{L}$ (left upper torso)]{\centering
% \setlength{\figureheight}{3cm}
% \setlength{\figurewidth}{3cm}
% \def\datapath{./figures/pgf/scatter_points_v3/m112/m112_llhf_K5}
% \input{./figures/pgf/scatter_points_v3/m112/m112_llhf_K5/m112_llhf_K5.tex}\hspace{-3mm}}
% \subfloat[$\mathsf{R}$ (right upper torso)]{\centering
% \setlength{\figureheight}{3cm}
% \setlength{\figurewidth}{3cm}
% \def\datapath{./figures/pgf/scatter_points_v3/m113/m113_llhf_K9}
% \input{./figures/pgf/scatter_points_v3/m113/m113_llhf_K9/m113_llhf_K9.tex}\hspace{-3mm}}
% \subfloat[$\mathsf{0}$ (reference)]{\centering
% \setlength{\figureheight}{3cm}
% \setlength{\figurewidth}{3cm}
% \def\datapath{./figures/pgf/scatter_points_v3/m132/m132_llhf_K5}
% \input{./figures/pgf/scatter_points_v3/m132/m132_llhf_K5/m132_llhf_K5.tex}}
% 
% \caption{Estimated scatter points using MLE.}
% \label{fig:scatter_points}
% \end{figure*}

\begin{figure}[t!]%\hspace{-5mm}%\centering
%\vspace{1mm}
\centering
\subfloat[$\mathsf{L}$ (left upper torso)]{\centering
\includegraphics{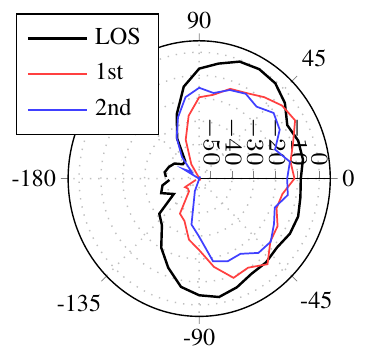}\label{fig:amp:L}\hspace{0mm}}
\subfloat[$\mathsf{C}$ (center upper torso)]{\centering
\includegraphics{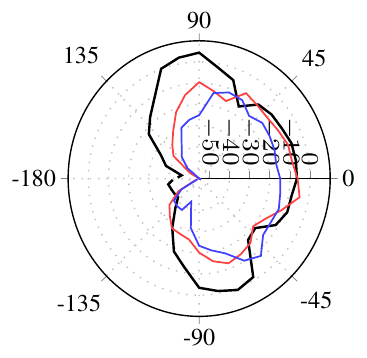}\label{fig:amp:C}}

\subfloat[$\mathsf{R}$ (right upper torso)]{\centering
\includegraphics{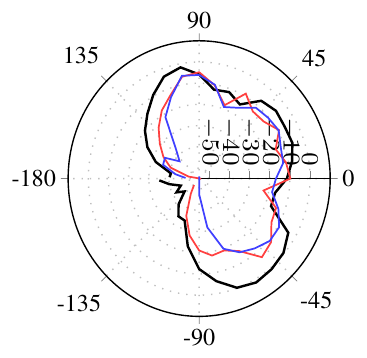}\label{fig:amp:R}\hspace{0mm}}
\subfloat[$\mathsf{0}$ (reference)]{\centering
\includegraphics{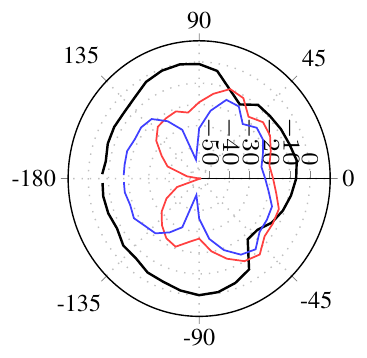}\label{fig:amp:0}}

\caption{Squared LOS magnitude (black) and squared average scattering coefficient magnitude of strongest (red) and second strongest (blue) average scattering coefficient magnitude plotted over angle of arrival $\varphi$. Quantities shown are averaged in 36 non-overlapping uniform angular sectors.
}
\label{fig:amp}
\end{figure}

Fig.~\ref{fig:amp} shows the squared LOS magnitude and squared scattering coefficient in dB as functions of the AOAs $\varphi_m$ of each measurement $\alpha_{m,j} \triangleq \alpha_j(\varphi_m)$ and $\beta_{m,j} \triangleq \beta_j(\varphi_m)$ respectively, with the $0^\circ$-direction indicated in Fig.~\ref{fig:scatter_points}. 
Positive angles are defined counter clockwise (see also Fig.~\ref{fig:scattering_environment}). 
For better visualization all quantities are averaged in 36 uniformly spaced sectors, and only the curves for the LOS and the scattering points with the largest and second largest average scattering coefficient magnitudes are shown.
Comparing the results for the on-body positions with the reference agent shows the strong influence of the human body when in between anchor and agent, i.e., when completely blocking the LOS in $180^{\circ}$ direction.
The notches at roughly $\pm 45^\circ$ in the reference antenna LOS magnitude in Fig.~\ref{fig:amp:0} are attributable to increased pulse dispersion in these directions, which is directly modeled by the larger scattering coefficient magnitudes, observable in Fig.~\ref{fig:amp:0}. % of the two strongest scattering points from these directions.
When comparing the on-body positions, the pattern of LOS magnitude and the scattering coefficient magnitude is of similar shape but slightly rotated. 
Note that this rotation is in the opposite direction w.r.t. the rotation of the agent position from the center upper torso.
By observing Fig.~\ref{fig:scatter_points:L} and \ref{fig:scatter_points:R} a possible explanation for this effect can be given by the following:
Moving the agent towards the left side of the upper torso and thus closer to the left hand/shoulder results in earlier shadowing by the left hand/shoulder, corresponding to earlier shadowing from positive AOAs.
The same reasoning holds for position $\mathsf{R}$.
For increased AOAs the scattering coefficient magnitudes decrease much faster than the LOS magnitude supporting the effect of lower pulse dispersion with AOAs close to $90^\circ$ before total blockage as observed in \cite{WildingPIMRC2020}. 

% \begin{figure}[t!]%\centering
% \subfloat[$\mathsf{C}$ (center upper torso)]{%\centering
% \setlength{\figureheight}{2cm}
% \setlength{\figurewidth}{5.5cm}
% \def\datapath{./figures/pgf/scatter_points_v3/m107/m107_alpha_K9}
% \input{./figures/pgf/scatter_points_v3/m107/m107_alpha_K9/m107_alpha_K9.tex}}
% 
% \subfloat[$\mathsf{L}$ (left upper torso)]{%\centering
% \setlength{\figureheight}{2cm}
% \setlength{\figurewidth}{5.5cm}
% \def\datapath{./figures/pgf/scatter_points_v3/m112/m112_alpha_K5}
% \input{./figures/pgf/scatter_points_v3/m112/m112_alpha_K5/m112_alpha_K5.tex}}
% 
% \subfloat[$\mathsf{R}$ (right upper torso)]{%\centering
% \setlength{\figureheight}{2cm}
% \setlength{\figurewidth}{5.5cm}
% \def\datapath{./figures/pgf/scatter_points_v3/m113/m113_alpha_K9}
% \input{./figures/pgf/scatter_points_v3/m113/m113_alpha_K9/m113_alpha_K9.tex}}
% 
% \subfloat[$\mathsf{0}$ (reference)]{%\centering
% \setlength{\figureheight}{2cm}
% \setlength{\figurewidth}{5.5cm}
% \def\datapath{./figures/pgf/scatter_points_v3/m132/m132_alpha_K5}
% \input{./figures/pgf/scatter_points_v3/m132/m132_alpha_K5/m132_alpha_K5.tex}}
% 
% \caption{Estimated scatter points using joint MLE.}
% \end{figure}

\begin{figure}[t!]\centering
% \hspace{-7mm}
\subfloat[average magnitude ratio (AMR)]{\hspace{0mm}%\centering
\includegraphics{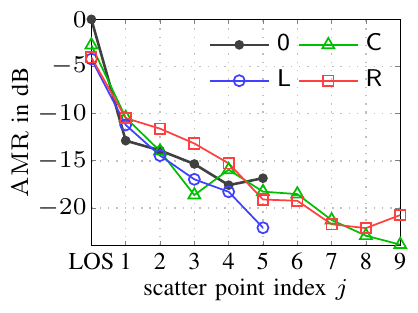}\label{fig:ratio:amr}\hspace{0mm}}
\subfloat[peak to average ratio (PAR)]{\hspace{0mm}%\centering
\includegraphics{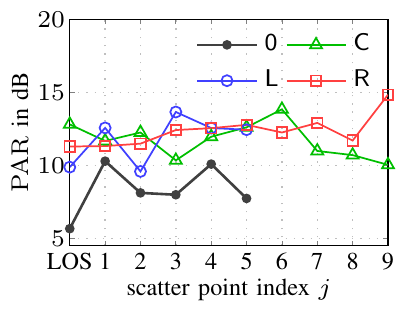}\label{fig:ratio:par}}

\caption{Ratio of average LOS magnitude or average scattering coefficient magnitude over the average LOS magnitude of the reference agent.}
\label{fig:ratio}
\end{figure}

Fig.~\ref{fig:ratio} shows the average magnitude ratio (AMR) and the peak to average ratio (PAR) in dB. %, defined below.
The former is the ratio of average LOS magnitude or average scattering coefficient magnitude to the maximum value of the LOS magnitude for the reference agent $\alpha^{(\mathsf{0})}_\mathrm{max} = \max_{m} |\alpha^{(\mathsf{0})}_m|$, i.e., 
\begin{align}
\mathrm{AMR}_{\text{LOS}}^{(\mathsf{X})} &= \frac{\bar{\alpha}^{(\mathsf{X})}}{\alpha^{(\mathsf{0})}_\mathrm{max}} &&\text{or}&\mathrm{AMR}^{(\mathsf{X})}_j = \frac{\bar{\beta}^{(\mathsf{X})}_j}{\alpha^{(\mathsf{0})}_\mathrm{max}}.
\end{align}
The latter is the ratio of maximum value to average value of LOS magnitude or scattering coefficient magnitude for each on-body position, i.e.,
\begin{align}
\mathrm{PAR}_{\text{LOS}}^{(\mathsf{X})} &= \frac{\max\limits_m |{\alpha}^{(\mathsf{X})}_{m}|}{\bar{\alpha}^{(\mathsf{X})}} &&\text{or}&\mathrm{PAR}^{(\mathsf{X})}_j=\frac{\max\limits_m |{\beta}^{(\mathsf{X})}_{m,j}|}{\bar{\beta}^{(\mathsf{X})}_j}
\end{align}
where $\mathsf{X}\in\{\mathsf{0},\mathsf{C},\mathsf{L},\mathsf{R}\}$ indicates the on-body position. 

The AMR curves in Fig.~\ref{fig:ratio:amr} show that for the reference agent $\mathsf{0}$ the scattering coefficients AMRs exhibit a steeper drop relative to the LOS AMR, while showing a similar trend as for the on-body positions. 
The AMR curves indicate that higher pulse dispersion can be expected with the agent at an on-body positions, due to the lower difference between LOS and scattering point AMRs. 
The plots also show the number of estimated scattering points for the different setups as $J_{\mathsf{0}}=J_{\mathsf{L}}=5$ and $J_{\mathsf{C}}=J_{\mathsf{R}}=9$. $J_{\mathsf{X}}$ is the number of scattering points estimated for on-body position $\mathsf{X}$. 

The PAR curves in Fig.~\ref{fig:ratio:par} show that even though the reference agent is far from an isotropic antenna, i.e., with $\mathrm{PAR}=0\,\mathrm{dB}$, it exhibits a generally lower PAR compared to the curves for the agent at different on-body positions. 
This leads to the conclusion of an increased directionality of the agent when at an on-body position. %a more isotropic amplitude pattern.
%Most notably, for the on-body positions $\mathsf{L}$ and $\mathsf{R}$, the PAR of the LOS is similar or slightly below that of all scattering points, implying a slightly larger field of view.

% TODO: add the information below into a table comparing the different on body positions
% \begin{itemize}
% \item peak to average amplitude ratio
% \item opening angle
% \item average amplitude in dB - plot all four in one plot indicating the number of points and how the avg amplitudes decay
% \end{itemize}
% 
% only use the strongest scattering points and indicate these in the plots, add plots with all estimated covariance matrices at once - include in fig 4
% 
% try plotting all in one plot, as polar plot, including the LOS and the strongest 2 scattering points, this could be included under fig 4

%%%%%%%%%%%%%%%%%%%%%%%%%%%%%%%%%%%%%%%%%%%%%%%%%%%%%%%%%%%%%%%%%%%%%%%%%%%%%%%%
%%%%%%%%%%%%%%%%%%%%%%%%%%%%%%%%%%%%%%%%%%%%%%%%%%%%%%%%%%%%%%%%%%%%%%%%%%%%%%%%
%%%%%%%%%%%%%%%%%%%%%%%%%%%%%%%%%%%%%%%%%%%%%%%%%%%%%%%%%%%%%%%%%%%%%%%%%%%%%%%%
\section{Conclusion}
\label{sec:conclusion}

In this paper we investigate joint modeling of an antenna and the human body, which add scattered components in addition to the line of sight path, as an extended antenna (EA).
% The extension of the model towards specular multipath propagation can be achieved in a straightforward manner, applying the same effects as observable to the LOS to all specular paths.
We briefly describe a stochastic EA model based on our previous work \cite{WildingPIMRC2020} and similar to \cite{SchubertTAP2013}, using a marked Poisson point process (MPPP) to model the delay dispersion introduced by antenna and human body.
Using measurements of a received ultrawideband signal with the agent positioned at different on-body positions on a static user, we calibrate the model for the different on-body positions, highlighting the differences of the on-body positions in comparison with the reference antenna. 
In all investigated setups the description using scattering points is feasible, with the different properties in terms of spatial spread of the scattering points and angular magnitude patterns of scattering points and LOS giving useful insights for modeling.
The results function as a starting point for a deeper analysis of the stochastic model parameter distributions.
% These differences in model parameters make it a sensible starting point for a deeper analysis of possible stochastic model parameter distributions.

Possible directions for future work are an in depth analysis of the model parameters in dynamic scenarios (based on different measurements of the same measurement campaign) and the application of this model in positioning and tracking algorithms with the goal of reducing the influence of the human body on the positioning quality.

%In future work, we intend to perform further validation of the model parameters found for the different setups but in dynamic user scenarios that were performed within the same measurement campaign.
%Following that, we will continue to investigate possible application of the model to positioning and tracking algorithms.

%%%%%%%%%%%%%%%%%%%%%%%%%%%%%%%%%%%%%%%%%%%%%%%%%%%%%%%%%%%%%%%%%%%%%%%%%%%%%%%%
%%%%%%%%%%%%%%%%%%%%%%%%%%%%%%%%%%%%%%%%%%%%%%%%%%%%%%%%%%%%%%%%%%%%%%%%%%%%%%%%
%%%%%%%%%%%%%%%%%%%%%%%%%%%%%%%%%%%%%%%%%%%%%%%%%%%%%%%%%%%%%%%%%%%%%%%%%%%%%%%%
% \appendix

%%%%%%%%%%%%%%%%%%%%%%%%%%%%%%%%%%%%%%%%%%%%%%%%%%%%%%%%%%%%%%%%%%%%%%%%%%%%%%%%
\bibliographystyle{IEEEtran}
\bibliography{IEEEabrv,./EuCAP}

\end{document}